\documentstyle[11pt,aaspp4]{article}
\begin{document}
%%\ \\
%%\hspace*{4truein} October 7, 1997\\[10truemm]

\title{CONSTRAINTS ON THE EXTRAGALACTIC INFRARED BACKGROUND FROM GAMMA-RAY
			OBSERVATIONS OF MKN 501}

\author{\bf Todor Stanev}
\affil{Bartol Research Institute, University of Delaware, Newark, DE 19716}

\author{\bf Alberto Franceschini}
\affil{Dipartimento di Astronomia - Universita' di Padova,
Vicolo Osservatorio 5, I-35122 Padova, Italy}

\begin{abstract}
 We use the new results of the HEGRA detector on the TeV $\gamma$--ray
 emission from MKN 501 to set upper limits on the energy 
 density of the cosmic infrared background (CIRB). Contrary to previous 
 interpretations of the $\gamma$--ray spectrum of MKN 421 as showing  
 an intergalactic absorption cutoff at 5 TeV, the
 observed spectrum of MKN 501 extends 
 beyond 10 TeV and appears to be unattenuated by $\gamma\gamma$ collisions
 with the low-energy CIRB photons. 
 The upper limits on the CIRB intensity -- derived both assuming an 
 {\it a priori} shape for the CIRB spectrum and without model-dependent
 assumptions -- are thus quite strong
 and come almost in conflict with the observational evaluations based on 
 deep surveys of
 extragalactic sources in the near- and mid-IR. If spectra at 
 TeV energies for extragalactic $gamma$-ray sources like this for MKN 501
 will be confirmed with 
 improved statistics, we may be forced to conclude that the process of 
 $\gamma\gamma$ interaction in the intergalactic space
 is more complex than expected and the average intergalactic magnetic 
field extremely weak ($B<10^{-11}\ G$). 
\end{abstract}

\keywords{infrared: general --- gamma rays: observations --- 
 scattering}

\section{Introduction}

 Cosmic history from the decoupling ($z=1500$) to the epoch of lighting
 of the first luminous sources (at redshifts $z\sim 3$ to 5) is one of
 the biggest unknowns of present--day observational cosmology.
 High redshifts and dust extinction during early active phases both
 degrade the energetic optical--UV photons emitted by massive stars,
 decaying particles, or more exotic energy sources, to the infrared
 wavelengths. A fundamental information on the total energy budget
 associated with astrophysical processes occurring at high redshifts is
 then provided by observations of the cosmic background at infrared
 wavelengths (CIRB). 

 Unfortunately, the infrared domain presents various levels of
 difficulty to the observational astronomer, because of the huge
 backgrounds from the Earth's atmosphere, the Interplanetary dust (IPD),
 and diffuse dust in the Milky Way, in addition to the background
 produced by the telescope itself. Also the sensitivity and stability
 of infrared detectors are far poorer than those used in the optical.
 Because of all this, the detection and characterization of the diffuse
 background flux of low-energy photons coming from primeval structures
 has been exceedingly difficult so far. 
 Even dedicated experiments exploiting cooled platforms outside the
 atmosphere, among which the most important is DIRBE on COBE
 (\cite{Hauser96}), have failed so far to detect significant
 signals from the CIRB above the intense foregrounds. The upper
 limits allowed by the foreground emission deconvolution of DIRBE
 maps are significantly higher than expectations over a substantial
 -- and crucial -- $\lambda$--range from a few to $\simeq 100$ $\mu$m. 
 The situation, at these wavelengths in particular, is not likely to
 improve in the future, until a mission flying to the outer Solar System
 will get rid of the fundamental limitation set by the IPD (both the
 scattered light and dust re--radiation). 

 Under these circumstances, a very interesting alternative to
 the direct detection of CIRB photons has been suggested by
 Stecker and de Jager (1993) soon after the discovery of high-energy
 photon fluxes coming from distant Blazars (with GRO, \cite{Hartman92}
%% Hartman 1992
 and with the Whipple Observatory, \cite{Punchetal92}).
%% Punch et al 1992).
 The idea is to infer the CIRB spectral intensity from combined GeV and
 TeV observations of a set of active galactic nuclei (AGN), by
 exploiting the $\gamma-\gamma$ interactions and pair production between
 the AGN high--energy photons and low--energy  background photons in the
 line--of--sight to the source. The interaction is expected to produce an
absorption feature, testable in principle, in the source TeV spectrum. 
Interesting limits have been discussed
%% by~\cite{SJ93,JSS94} and~\cite{DweckSlavin94},
 by Stecker and de Jager (1993), de Jager, Stecker and Salamon (1994),
 and Dwek and Slavin (1994), {\it all based on TeV observations of the Blazar
 MKN 421}. 

 To summarize, the best current upper limits on the CIRB in
 the spectral range from 10 to 40 micron are those reported
%% by~\cite{JSS94},
 by de Jager, Stecker \&  Salamon (1994), with a $2\sigma$ upper
 limit of $\lambda I_\lambda < 2\ 10^{-8}\ W/m^2/sr$. In the same
 large waveband interval, marginal detections, at  levels of
 $2\ 10^{-9} <\lambda I_\lambda < 2\ 10^{-8} W/m^2/sr$ were
 reported by  de Jager, Stecker \&  Salamon (1994)
 and Dwek \& Slavin (1994). At shorter wavelengths, 1 to 10 $\mu$m,
 upper limits have been obtained by
%%~\cite{SJ93,Stecker96} and~\cite{DweckSlavin94}.
 Stecker \& de Jager (1993), Stecker (1996), Dwek \& Slavin (1994)
 and Biller et al. (1995).
 The most conservative bounds, accounting for the precise spectral 
 shape of the CIRB, given by Dwek \& Slavin (1994,
$\lambda I_\lambda < 10^{-7}\ W/m^2/sr$), still keep a substantial factor
 ($>$10) higher than the expected contribution of known sources.
 
While all previous analyses relied uniquely on TeV observations of the blazar 
MKN 421, we exploit here a new high--quality dataset of TeV gamma--ray
 observations by HEGRA of MKN 501 during a state of high activity
 (Aharonian et al 1997a) to further constrain the intensity of the
 diffuse IR background. 

 Two sets of constraints on the CIRB spectral intensity are derived
 in Section 2. One is based on the assumption that the background
 spectrum is dominated by the contribution of distant galaxies,
 and thus reflects the average galactic IR spectrum.
 The other constraint is free of model--dependent assumptions 
 and treats the CIRB as a combination of twelve bins wherein
 the background spectrum is flat (in $\lambda I_\lambda$) with
 independent arbitrary normalizations. Extremely tight constraints
 on the CIRB ensue from this analysis, reflecting the missing
 evidence of any absorption in the TeV spectrum of MKN 501.
 A discussion is given in Section 3 
 in terms of a low average past emissivity both of galaxies and of 
 primeval energy sources. 
 Indeed the limits are so severe that they begin to 
 conflict with the integrated IR flux of distant galaxies, recently detected 
 in large numbers by ground-based and space observatories. 
 We finally emphasize the alternative possibility that the interaction 
 of high-energy gamma-rays with low-energy photons
 is more complex than previously supposed.
 $H_0=75\ km/s/Mpc$ is assumed throughout the paper.

\section{Data analysis and results}

 We use the data set taken in March and April 1997 with  the HEGRA 
 stereoscopic system of four imaging Cherenkov telescopes (\cite{Hermann95}). 
 At that time MKN 501 was in an extremely high state, the far brightest 
gamma-ray source in the sky. The flare 
 was observed by all TeV $\gamma$--ray observatories in the 
 Northern hemisphere (\cite{Breslinetal97}). The high gamma--ray flux
 allowed the HEGRA observatory to measure the gamma--ray energy spectrum
 in short time intervals. The spectrum measured between March 15 to 20
 covers 1.2 orders of magnitude in photon energy, from 0.8 to 12.6 TeV,
 divided into twelve logarithmically spaced energy bins. The photon number
 spectrum at TeV $\gamma$--ray energies is consistent with a power law 
 of differential spectral index $\alpha$ = 2.49$\pm$0.11. This spectrum
 is shown in Figure~1 (data points) together with the allowed range of
 best--fitting power--laws derived by the observational team (shaded region). 
Similar power-law spectra for MKN 501 have been observed by various
other groups (see e.g. Protheroe et al. 1997).

 The main difficulty with the derivation of the CIRB spectral
 intensity from TeV $\gamma$--ray absorption is the lack of
 knowledge of the production spectrum, which may deviate from
 a pure power-law and/or show absorption at the source. Spectra
 measured within short time intervals, during which the source's activity state
is not likely to vary, are very
 valuable in this respect, although there is not an evidence for strong
 spectral variability of the MKN501 emission(\cite{aharon97b}).
For this reason we confined our analysis to a spectrum based on data collected
between March 15 to 20, rather than the one using the whole March-April
database in the updated version of Aharonian et al (1997a).

 As a first step in understanding the observed spectrum, we attempted
 to derive the allowed range of differential spectral indices $\alpha$
 at the source and the CIRB absorption by fitting the observed 
$\gamma$--ray spectrum with a variety of assumptions for $\alpha$ and
 for the intensity $I_{IR}$ of the extragalactic IR background. We
 first assumed that the source spectrum in the observed range is a pure
 power-law and that the low-energy background is dominated by the
 integrated contribution of distant galaxies. In such a case, the shape
 of the CIRB is well constrained, and should reflect 
the average galactic spectrum,
 which has a minimum at $\lambda \sim 10 \mu m$ corresponding
 to the intersection of the stellar with the dust emission component.
 We refer here to the detailed spectral shape estimated by
 Franceschini et al (1991). 

 Then the fit was performed in a two parameter space consisting of the
 source gamma-ray spectral index $\alpha$ and the CIRB intensity
 $I_{IR}$ normalized to the model spectrum $I_0(\lambda$) by 
 Franceschini et al (1991). For a given spectral index $\alpha$ the
 absorption due to the 
CIRB was first calculated and the resulting spectrum was
 then normalized to the detected $\gamma$--ray flux above 1 TeV. 
 This procedure enabled us to 
 avoid introducing a third parameter, the source spectrum normalization. 
 Table~1 shows the optical depth assuming $I_{IR}/I_0$ = 1, for all twelve
 experimental energy bins. 

 Figure~2 summarizes the results of fitting the observed MKN 501 spectrum. 
 It plots contours of the $\chi^2$ 2D distribution in the parameter
 plane, including the unphysical region corresponding to negative 
$I_{IR}/I_0$ values, where 
 gamma--rays are `created' in collisions with CIRB photons. 
The darker an area is the better the fit is (except for fits with 
$\chi^2\,<$ 1.45, corresponding to the blank inner region). 
The formal best fit ($\chi^2_{\nu}$ = 1.41 p.d.f.
 for 10 degrees of freedom, confidence level 0.16) is in the ``unphysical''
 region: $\alpha$ = 2.52, $I_{IR}/I_0$ = -0.3. The best fit in the
``physical'' region occurs at $I_{IR}/I_0)$ = 0 (no-absorption,
$\chi^2_{\nu}$ = 1.42 p.d.f.) 
and yields $\alpha$ = 2.48, in agreement with the
 result of the experimental group. Flatter TeV $\gamma$--ray source
 spectra allow for higher intergalactic absorption:
 a value for the spectral index of $\alpha=2.16$, corresponding to
 $I_{IR}/I_0$ = 1.70, is the hardest spectrum with $\chi^2_{\nu}\,<$ 1.80
(corresponding to the 95\% confidence limit).
%evel of a fit with $\chi^2$ = 1.80 is smaller than
%that of the best fit by a factor of three. 
The best-fit spectra corresponding to $I_{IR}/I_0$=0 and the flattest 
one assuming $I_{IR}/I_0$ = 1.7 are also shown in Fig.~1.

 It is important to understand that the fit quality, and correspondingly
 the limits on the CIRB, depend crucially on the manner in which the
 `theoretical' power-law fluxes are normalized in the fitting procedure.
 The normalization to the observed $\gamma$--ray flux above 1 TeV causes
 very different spectral indices to fit the observations equally well and
 generates a broad $\chi^2$ valley in Fig.2. This would change significantly
 if the observed $\gamma$--ray energy range were extended on either side.
 An observation at $E_\gamma\,\geq$ 300 GeV (the threshold energy of the
 Whipple telescope) would improve the overall normalization and allow us
 to distinguish better between different spectral indices `at production'.
 An extension to higher $\gamma$--ray energy, hence larger optical
 depths (see Table~1) would make easier the detection of any absorption
 by the CIRB.

%% One could use the fits within the parameter space shown in Fig.~2 
%% to set more specific upper limits on the CIRB energy density as a
%% function of the IR photon wavelength, still under the assumption
%% that our adopted theoretical shape of the CIRB spectrum is correct
%% to the first order. For each $\gamma$--ray energy bins in Fig. 1,
%% the maximum absorption is estimated
%% as ($F_\gamma$ - 1.64$\delta F_\gamma$)/$F_{prod}$, where $F_\gamma$
%% and $\delta F_\gamma$ are the experimentally measured $\gamma$--flux
%% and its error. $F_{prod}$ is the unattenuated value  that corresponds
%% to a spectral index $\alpha$ normalized to the total detected flux
%% after absorption (the factor 1.64 converts from 1$\sigma$ to a 90\% 
%% confidence limit). 
%
%Figure~3 shows
% the highest values for $I_{IR}$ allowed by the non--absorption of the
% TeV $\gamma$--rays for any of the primary spectra indices (2.16 to 2.68)
% that fit the spectrum better than 1$\sigma$ in the physical region
% shown in Figure~2. For each $\gamma$--ray energy bin we used  the
% wavelength interval that contribute 90\% of the optical depth.
 
 We finally attempted to obtain model--independent limits on the CIRB,
 with no a-priori guess at the background spectrum.  To compute them
 we conservatively  assumed that the high-energy source spectrum is
 the flattest allowed by the fits of Fig.~1, i.e. $\alpha$ = 2.16.
 We normalize the spectrum at the source to be $F_\gamma + 1.64
 \times \delta F_\gamma$ in the first experimental bin. This
 normalization factor is a source  of uncertainty. It is partially
 justified by the fact that MKN 501 is close enough (136 Mpc) compared
 to the optical depth (600 Mpc for $ \lambda I_\lambda = 7.64\ 10^{-9}
 \ W\ m^{-2}\ sr^{-1}$) that the content of this bin would be absorbed
 only by 20\%.

The absorption of $\gamma$--rays of energy $E$ TeV is due to IR
photons within a certain wavelength interval around the maximum
absorption at $\epsilon_{max}$ = $2 (m_e c^2)^2/E_\gamma$
($\lambda_{max} = 1.24/(\epsilon_{max}, eV)\; \mu$m). 
 We have then assumed that the absorption of any given $\gamma$--ray energy
 bin is caused by CIRB with flat $\lambda I(\lambda)$ spectrum. The envelope
 of all upper limits to the CIRB intensity
obtained in this way is shown as a histogram in Fig.~3. 
 The limits are assigned within wavelength intervals that contribute 
90\% of the optical depth for that bin.
%% that at any given $\gamma$-ray energy bin the
%% absorption is contributed by a CIRB background with flat 
%% $\lambda I(\lambda)$ spectrum, the IR wavelength interval
%% corresponding to all photons contributing 90\% of the optical depth
%% for any $\gamma$-ray bin. 
%% The envelope of all upper limits obtained in this way is shown as a
%% histogram in Fig.~3. 
% For each CIRB wavelength we show the highest limit coming from any of
% the $\gamma$--ray energy bins. 
 As we see, the limits become less stringent at
 longer wavelengths. The limit in the 3.4 $\mu$m to 24 $\mu$m range
 comes from one single energy bin (5.0 to 6.3 TeV) where  a particularly
 large $\gamma$--ray flux was measured. 

 {\it  Note that the obtained upper limits
%% obtained in such a way
 fall already very close to recent direct evaluations of the IR background
 based on deep IR surveys (\cite{Franceschini97})}. As shown in
 Fig. 3, the upper limits heavily rely on the error in the TeV
 flux measurement and are less stringent for $\lambda >$ 10 $\mu$m, where
 the TeV $\gamma$--ray statistics is not as good. 

\section{Discussion}
  
 In either case, both assuming the CIRB shape and relaxing it,
 the constraints on the CIRB intensity appear dramatic. Essentially
 the MKN 501 gamma-ray spectrum does not display the expected effect
 of absorption, it rather shows a simple $E^{-2.5}$ power-law
 spectrum. 

 How reliable are these limits in view of the possible systematic
 errors of $\sim$25\% in the energy estimates of the HEGRA
 telescopes (\cite{aharon97c})? This is very easy to estimate in
 the case of a flat binned $\lambda I_\lambda$ CIRB spectrum. 
 The limits in Figure 3 would
 move upward and towards shorter wavelengths with the fractional
 amount of energy overestimate.  Similar relaxation would occur
 also in the case of a more specialized 
model CIRB spectrum. One could use the
 optical depths from Table~1 to estimate the amount of relaxation. 
 Similarly, a higher normalization of the source spectrum 
would relax the  model--independent limits by the ratio
 of the two normalizations.

 Is this featureless spectrum of MKN 501 inconsistent with that
 observed for MKN 421, which is almost at the same distance? The
 latter has been interpreted by some authors (e.g. \cite{Stecker96})
 as showing a turnover at $\epsilon \simeq 3-5\ TeV$,
 which was attributed to $\gamma\gamma$ absorption with the CIRB. 
 In fact, new observations of this source during a high activity state
 do not appear to confirm the presence of absorption (Krennrich et al. 1997),
 and show significant counting rate above 5 TeV. So, at least during this
 high state, MKN 421 seems to show a power law spectrum similar
 to the spectrum discussed here for MKN 501.

 The constraints on the CIRB intensity from TeV observations of
 MKN 501 start to approach the "measured" lower limits at 2.2,
 6.7 and 15 $\mu$m given by the integrated emission of  galaxies
 already resolved in deep integrations at those wavelengths.
 Deep surveys have been performed from ground in the K-band 
 and from space by the mid--IR camera (ISOCAM, see~\cite{Cesarskyetal96})
 on the ISO satellite in the two latter bands. A summary of
 these "direct" determinations of the galaxy  contribution to the
 CIRB, and a discussion of the related uncertainties, are given by
 Franceschini et al (1997) and Oliver et al. (1997). 
 In any case, the CIRB cannot be lower than reported at these three 
 wavelengths. 

 Few possibilities are left. The first one is that the CIRB is very 
 close to the limits
 allowed by the gamma--ray spectrum of MKN 501 observed by 
 Aharonian et al. (1997a). This would imply a very strong constraint
 on any signals unrelated to the emission of distant galaxies
(see e.g. \cite{RRC88}, for a review).

% The constraint seems particularly impressive in the near--IR,
% at 1 to 2 micron: the residual background, after subtraction of
% the contribution of galaxies, must be much less than 
% $\lambda I(\lambda) = 10^{-8}\ W/m^2/sr$.
% Any significant emission at high redshifts ($z$=1 to 100) from
% astrophysical sources other than galaxies (see \cite{RRC88},
% for a review) would be ruled out if the limits presented here
% are confirmed by similar multi--TeV $\gamma$--ray observations.

 But, in view of the fact that the same power-law spectral shape as
 show in Fig. 1 for MKN 501 has been confirmed by later integrations
 on this source (Aharonian et al. 1997a; Protheroe et al. 1997), 
that apparently a similar
 shape is also suggested for MKN 421, and because an appreciable CIRB
 flux has already been detected and resolved into discrete sources,
 we find more likely that {\it we have to revise our concepts about
 the propagation of TeV gamma-rays into the intergalactic space, and that 
 something complicates the process}.

 The question is: why the propagation of TeV gamma-rays in intergalactic 
 space should 
not produce the expected absorption in high energy spectra 
 of distant sources?  A possible solution could be that part of the source
 spectrum is regenerated in $\gamma$--ray cascading (pair production
 + Inverse Compton). In such a cascading process, the $\gamma$--ray
 spectrum of the source is depleted around the region of maximum
 absorption. If the $\gamma$--ray emission of MKN 501 at the 
 source extends above $10^{14}$ eV, the spectrum would be depleted
 in collisions with microwave background photons. The $e^+e^-$ pairs
 generated on the microwave background would Inverse Compton scatter
 on the microwave background to regenerate photons of lower (TeV)
 energy, thus generating `bumps' on a power-law production spectrum
 (\cite{ProthSta93}). The resulting $\gamma$--ray  spectrum may then
 appear unattenuated at observation. This would however require
 not only a $\gamma$--ray spectrum extending to very high energy,
 but also  a
 very low ($\sim 10^{-11}$ Gauss) value for the extragalactic magnetic
 field in the direction of MKN501. Otherwise the $e^+e^-$ pairs would
 deflect in the magnetic field and form a halo around the source, well
 outside of the angular resolution of the HEGRA detector. 

 A good deal of constraints useful to disentangle between these two
 possibilities are soon expected by improved observations of the
 MKN501 outburst (which has been observed by the Whipple and CAT Cherenkov
 telescopes, \cite{Breslinetal97}) with different energy thresholds
 and wavelength bands and by refined forthcoming data on MKN 421. \\[2truemm]
 {\bf Acknowledgements.} The authors appreciate the contribution of 
 an anonymous referee to the improvement of the paper. TS thanks 
 J. Buckley and T.K.~Gaisser for the careful reading of the manuscript
 and E. Dwek for comments. The research
 of TS is funded in part by NASA grant NAG5--5106.
\vfill\eject

\vfill\eject
\centerline{FIGURE CAPTION}

\figcaption{ The energy spectrum of the TeV $\gamma$--rays
 from MKN 501 observed by the HEGRA detector, together with
 its power--law spectral fits (lightly shaded area). The
 solid line shows our best fit in the physical region (see text).
 The dashed line shows the fit with $\alpha$ = 2.16 and
 $I_{IR}/I_0$ = 1.70.
}

\figcaption{ The range of spectral indices and CIRB energy
 densities that allow for 90\% confidence fits of the observed
 TeV $\gamma$--ray spectrum of MKN 501, assuming a theoretical
 shape for CIRB -- see text. The best formal fit, shown with a
 cross, is in the unphysical region. The $\chi^2$ scale is given
 in the upper right corner.
}

\figcaption{ Upper limits for the CIRB density derived from the TeV 
 $\gamma$--ray spectrum of MKN 501: a) assuming the theoretical shape
 shown with a thin line; b) assuming flat $\lambda I_\lambda$. The thick
 line shows the estimate of Franceschini et al (1991) and the three
 data points are from Franceschini et al (1997). These are three direct
 evaluations of the CIRB spectral intensity due to faint galaxies 
 at $\lambda=2.2$, 6.7 and 15 $\mu m$, based on galaxy models fitting
 deep counts performed at these wavelengths.
}
\newpage

\begin{table*}
\tablenum{1}
\caption{ Optical depths and CIRB wavelength ranges responsible
 for the absorption of TeV gamma rays by MKN 501 for a distance
 of 136 Mpc ($H_0$ = 75 km/s/Mpc). Column 1 shows the $\gamma$--ray
 energy range, columns 2, 3 \& 4 give the optical depth, CIRB
 wavelength of maximum absorption and the wavelength range responsible
 for 90\% of the optical depth for the model of Franceschini et al
 (1991). Columns 5, 6 \& 7 give the same quantities for constant
 $\lambda I_\lambda = 7.64\ 10^{-9} \ W\ m^{-2}\ sr^{-1}$.
 \label{tbl1}}
\begin{center}
\begin{tabular}{crrcrrc}
\tableline
 $E_\gamma$ & \multicolumn{3}{c}{model $\lambda I_\lambda$} &
  \multicolumn{3}{c}{$\lambda I_\lambda$=const} \\
 (TeV) & $\tau_{\gamma\gamma}$ & $\lambda_{max}$ & $\lambda_{90\%}$ &
   $\tau_{\gamma\gamma}$ & $\lambda_{max}$ & $\lambda_{90\%}$ \\
  & & ($\mu$m) & ($\mu$m) &  & ($\mu$m) & ($\mu$m) \\ 
\tableline
  0.79 -- 1.00 & 0.25 &  1.7 & 0.53 -- 3.8 & 0.23 & 2.1  & 0.53 -- 3.8 \\
  1.00 -- 1.26 & 0.29 &  2.0 & 0.60 -- 4.8 & 0.28 & 2.7  & 0.67 -- 4.8 \\
  1.26 -- 1.59 & 0.32 &  2.1 & 0.67 -- 6.0 & 0.36 & 3.4  & 0.84 -- 6.0 \\
  1.58 -- 1.99 & 0.33 &  2.2 & 0.75 -- 7.5 & 0.45 & 4.2  & 1.1 -- 7.5 \\
  2.00 -- 2.51 & 0.34 &  2.4 & 0.84 -- 9.5 & 0.57 & 5.3  & 1.3 -- 9.5 \\
  2.51 -- 3.16 & 0.35 &  2.5 & 0.95 -- 12. & 0.71 & 6.7  & 1.7 -- 12. \\
  3.16 -- 3.98 & 0.37 &  3.0 & 1.1 -- 15.  & 0.86 & 8.4  & 2.1 -- 15. \\
  3.98 -- 5.01 & 0.41 & 13.4 & 1.2 -- 19.  & 1.13 & 10.9 & 2.7 -- 19. \\
  5.01 -- 6.31 & 0.48 & 15.0 & 1.5 -- 24.  & 1.43 & 13.4 & 3.4 -- 24. \\
  6.31 -- 7.94 & 0.59 & 16.8 & 1.9 -- 30.  & 1.80 & 16.8 & 4.2 -- 30. \\
  7.94 -- 10.0 & 0.76 & 21.2 & 2.4 -- 38.  & 2.27 & 21.2 & 5.3 -- 38. \\
 10.00 -- 12.6 & 1.00 & 30.0 & 3.4 -- 48.  & 2.85 & 26.7 & 6.7 -- 48. \\
\tableline
\end{tabular}
\end{center}
\end{table*}
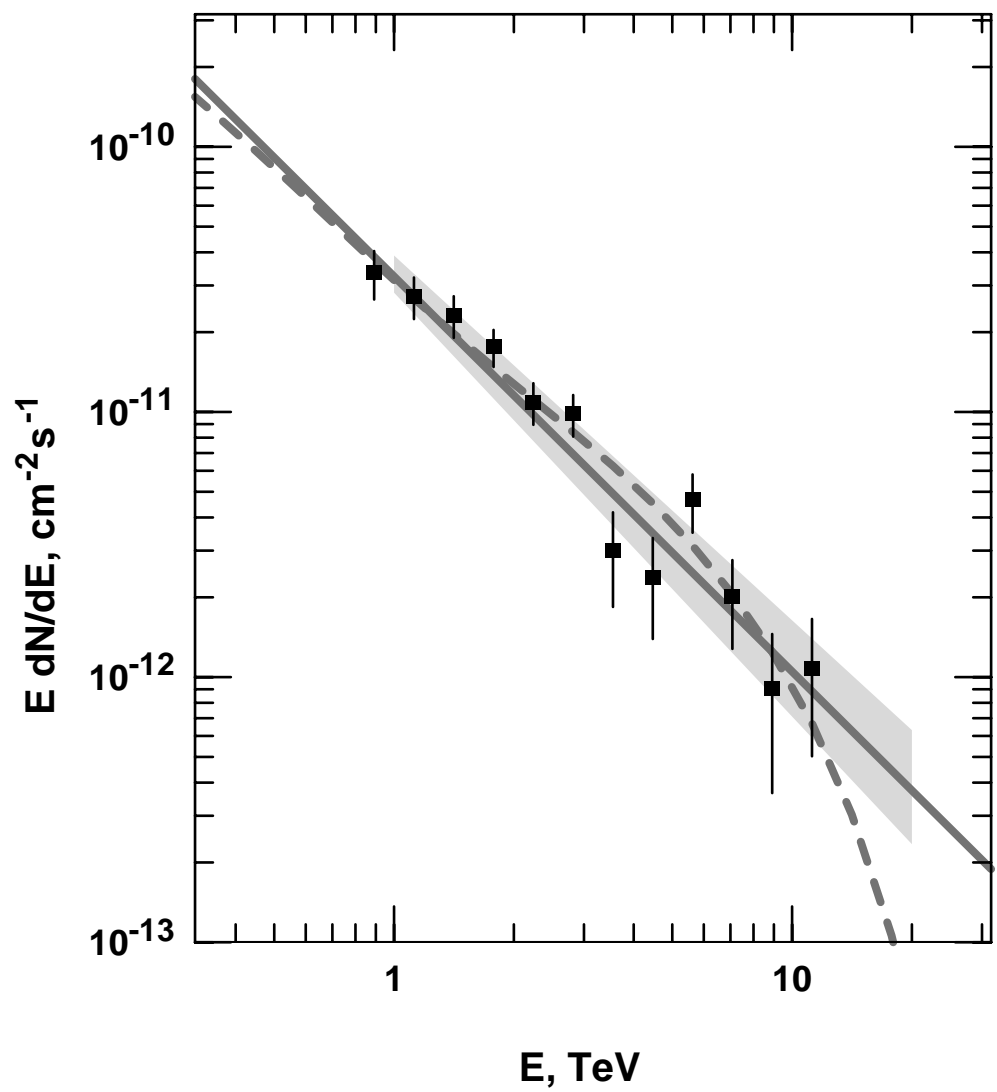
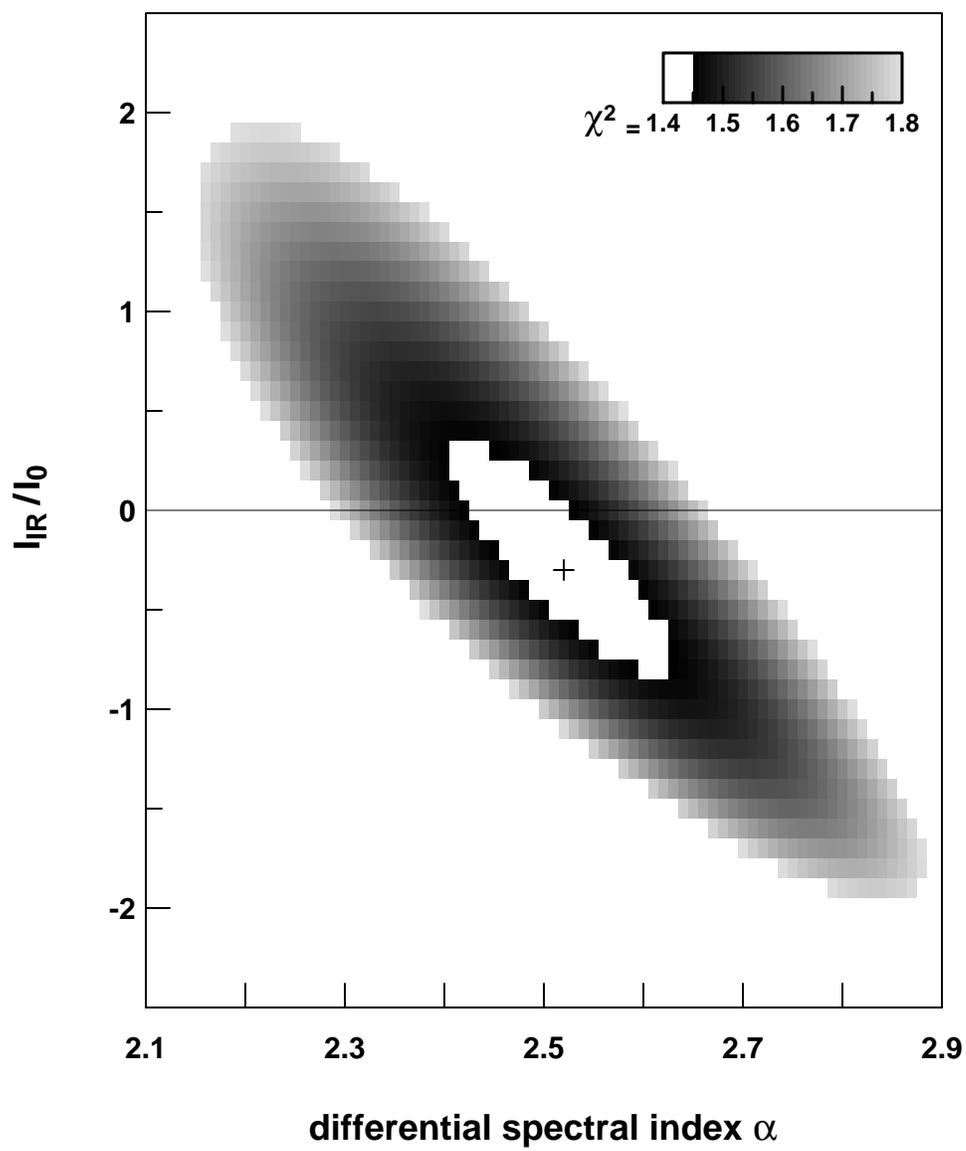
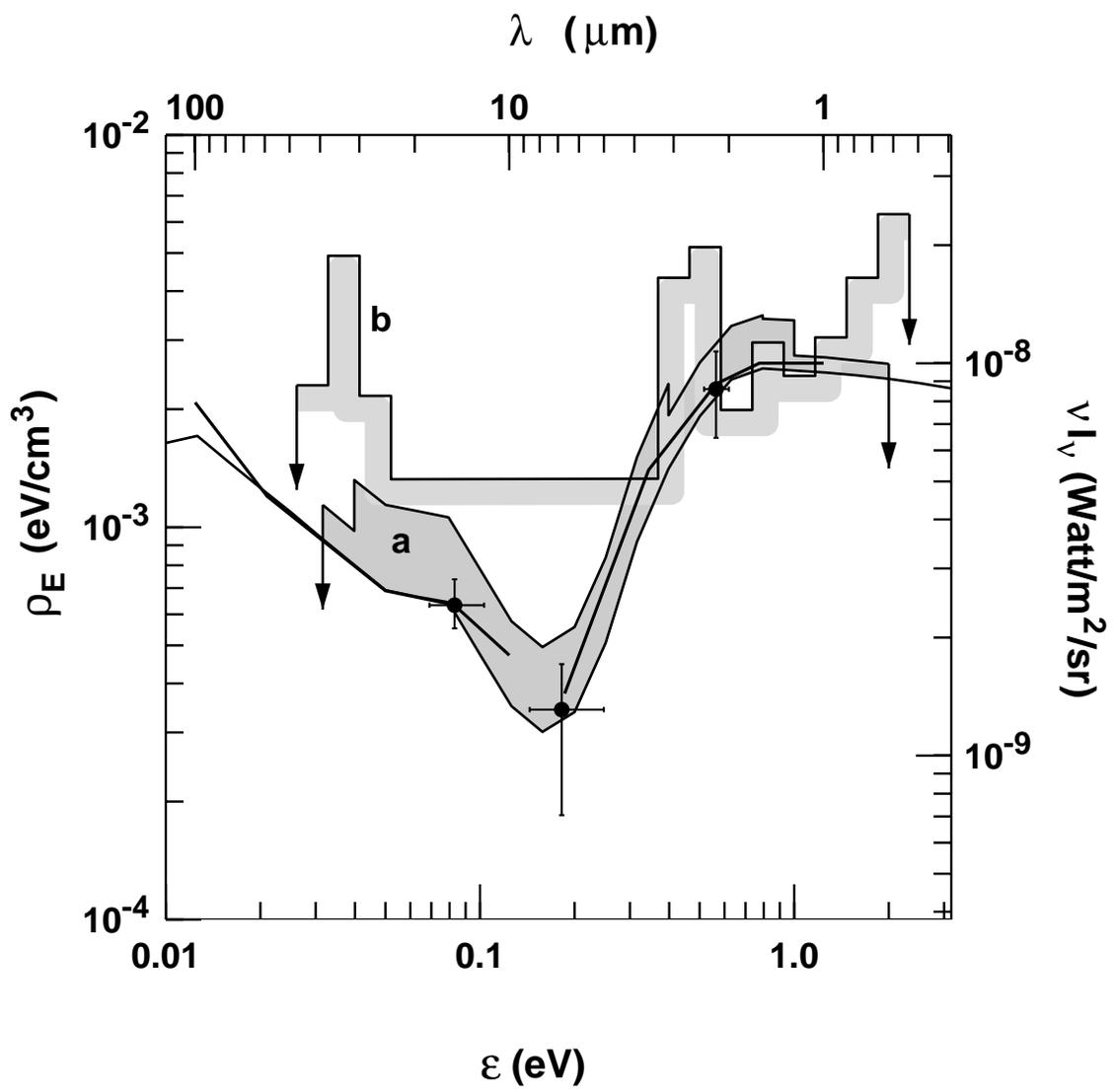
\end{document}